# Sensus Pond: Exploring Water as Sensing Medium for More-than-Human Observation

Wu, Kuan-Ju*[a]; Hu, Youyang*[a]; Chou, Chiaochi[b]; Kakehi, Yasuaki[a]

[a] The University of Tokyo, Tokyo, Japan
[b] National Tsing Hua University, Hsinchu, Taiwan
* kuanju@xlab.iii.u-tokyo.ac.jp
* These authors contributed equally

We introduce *Sensus Pond*, an interactive system that reconfigures water not merely as a static natural element but as an active sensing medium for registering more-than-human traces. In response to the limitations of anthropocentric approaches in interaction design, we propose a methodological framework of *observation without translation*—resisting the tendency to stabilize, decode, or humanize nonhuman presence. Drawing from critical theories of more-than-human design and ecological entanglement, Sensus Pond is a materially embedded and site-specific system that employs Swept Frequency Capacitive Sensing (SFCS) and an Artificial Neural Network (ANN) to detect ephemeral interactions between the pond's surface and surrounding life forms. Rather than classifying or interpreting these events, the system visualizes temporal accumulations of overlapping traces—producing a layered archive of spatial and temporal entanglements. This approach emphasizes *attunement over control*, shifting the designer's role from interpreter to facilitator of open-ended, multispecies encounters. Sensus Pond invites reflection not only on what is sensed, but on how design itself can remain responsive to ambiguity, contingency, and the aesthetics of shared ecological life.

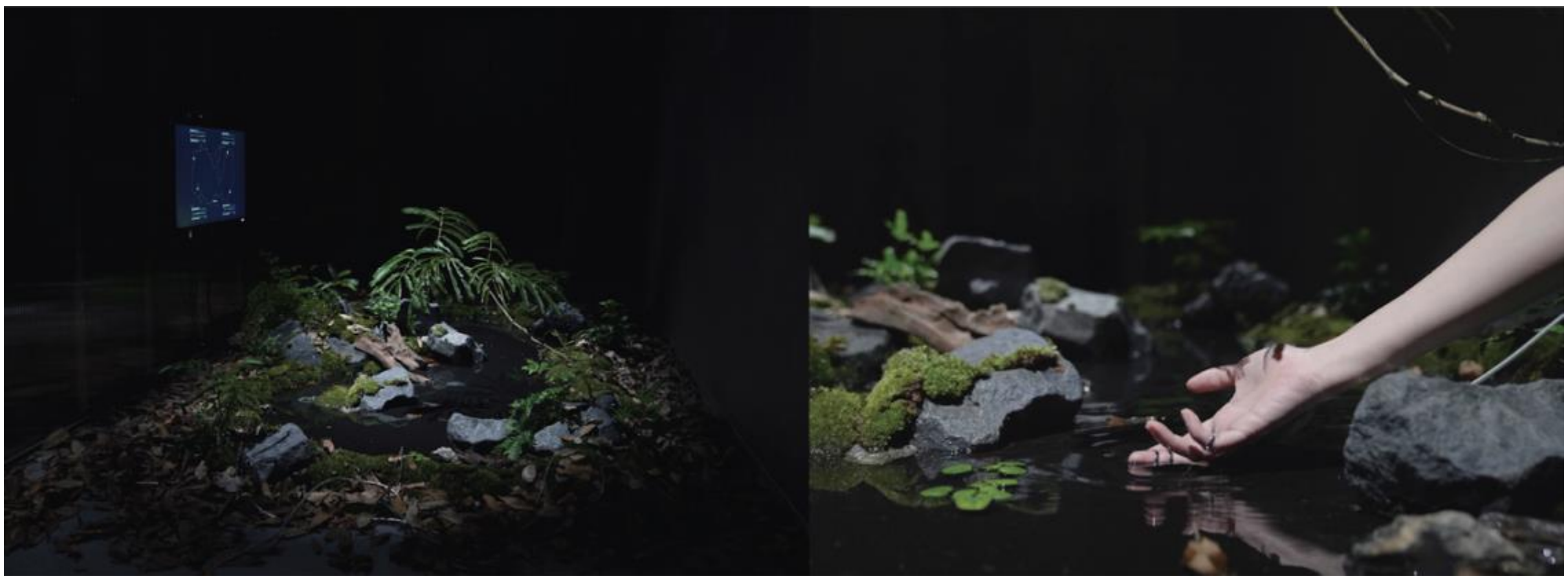

*Figure 1. Sensus Pond is an interactive system that integrates water as a sensing medium for more-than-human observation. The left image illustrates the Sensus Pond within an indoor setting, and the right image depicts a human participant interacting with the water through physical contact.*

# 1 Introduction

Design research has shifted from prioritizing human needs and usability to engaging with the entangled realities of our shared world. While human-centered frameworks once democratized technology, they also obscured the interdependencies linking human and nonhuman life. As planetary crises intensify, the limits of anthropocentric design are clear, prompting a reimagining of design as ecological attunement and multispecies care.

This shift redefines what counts as interaction and participation when agency is distributed across species and materials. It invites sensing to move from measurement toward attunement—using technologies of detection to stay with ambiguity rather than resolve it. Within this context, our project investigates how interactive sensing systems might register traces and rhythms of presence not as data to decode, but as material expressions that sustain openness, partiality, and coexistence.

## 1.1 Theoretical reorientations: from dualism to entanglement

Efforts to reconnect with the natural world demand a theoretical shift that challenges the anthropocentrism and nature–culture divide of Western modernism (Wolfe, 2001). Feminist scholars in science and technology studies have proposed relational and material understandings of existence. Barad's (2003) notion of intra-action reframes being as emergent through relation, where entities take form within entanglement rather than before it. Haraway (2016) extends this ethically, urging us to stay with the trouble—to live with complexity, opacity, and shared responsibility in a more-than-human world.

Together, these perspectives reorient design toward attunement rather than control, emphasizing responsiveness to evolving relations. Meaning and agency arise through mutual shaping among bodies, materials, and environments, offering a foundation for approaches that engage relations materially and reflectively rather than abstractly.

In design research, these ideas inform explorations attentive to nonhuman agencies and rhythms through situated material practice (Giaccardi et al., 2025; Rosén et al., 2022). Our inquiry flows from this current of thought, exploring how design can engage with subtle relations that emerge when we observe with openness rather than control.

## 1.2 Critical reflections on more-than-human design

These theoretical insights have shaped a new design paradigm that reconsiders how we engage with nonhuman forms—materially, ethically, and methodologically. Recent projects have explored sensing with plants (Chang et al., 2022; Hu et al., 2024), fungi and protists (Liu et al., 2018; Lu & Lopes, 2022; Ofer & Alistar, 2024), microbes (Pataranutaporn et al., 2020; Zhou et al., 2024; Bell et al., 2023), and insects (Ikeya et al., 2023; Wakkary et al., 2023), among others. Across this growing body of work, some translate biological data into emotional metaphors (e.g., "happy/sad plant"), offering accessible ways to relate to ecological processes while exploring how affect can mediate more-than-human engagement. In contrast, Wakkary et al. (2021) embrace ambiguity, crafting interactions that resist closure and foreground the limits of human understanding.

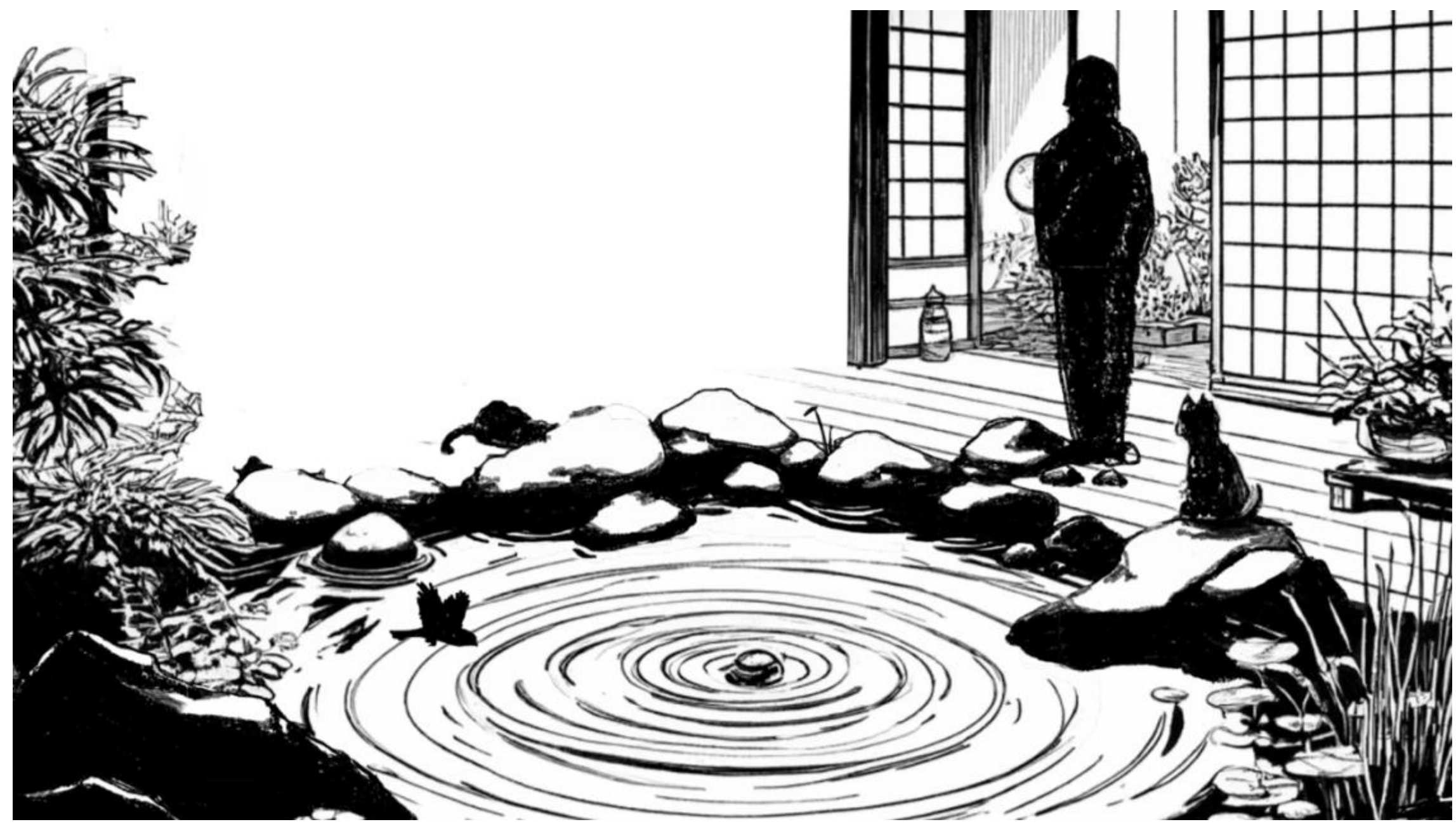

*Figure2. Concept sketch of the Sensus Pond environment.*

These diverse approaches reveal both the promise and complexity of sensing-based engagement. Translating nonhuman phenomena into human contexts can foster empathy and participation, yet it also requires attentiveness to difference—the quality that makes such encounters meaningful. The issue lies not in representation itself, but in the human impulse to render the unfamiliar immediately knowable. As Hui (2021) cautions, "we have yet to reflect on how, for our epoch, the Unknown can be demystified and de-anthropomorphized while remaining effective." Designing for ambiguity, then, becomes a way to sustain the unknown—to engage without collapsing difference into human terms, allowing nonhuman presences to appear on their own conditions.

## 1.3 Tracing engagement with pond

In response, we propose an alternative mode of engagement—one that resists decoding or symbolic translation. We turn to a pond: a quiet surface where relation becomes perceptible. Why a pond? The pond gathers this sensibility: it is a site of multispecies overlap and liminality —a reflective field where relations appear and dissipate(Wakkary, 2025). Across its surface, insects, birds, cats, and humans leave subtle disturbances—ripples, shadows, and conductive fluctuations that register contact yet elude interpretation. These are not messages but ephemeral signs: ambiguous, shifting, and open-ended (Offenhuber, 2020).

Here, tracing is both a material practice and a way of thinking—an orientation toward what is emergent, partial, and unresolved. It situates design within processes of entanglement rather than representation. Traces, as Rosner et al. (2013) describe, record relation—trajectories formed through movement and touch. The pond's ripples and microcurrents compose a living archive of such encounters, recording without inscription. To trace is not to capture events but to follow how matter remembers through motion. It is also a strategy for orienting toward nonhuman perspectives, cultivating awareness through noticing and care (Ofer & Alistar, 2024).

In Sensus Pond, tracing becomes a practice of attunement. It encourages slowness, uncertainty, and attention to subtle change. Each fluctuation in the water acts as both record and prompt—to consider what we follow, how we frame, and what we allow to remain unresolved. The pond operates as both medium and method: a material field where design unfolds through coexistence rather than control, and where observation turns into shared presence.

### 1.4 Observation without translation

Building on tracing as a stance of attunement, we explore how observation might unfold without translation. To observe the pond is not to extract meaning but to remain within its shifting conditions. In Sensus Pond, sensing becomes a way of coexisting rather than measuring. The capacitive traces we collect are not data to decode but gestures of contact—moments when surfaces meet and briefly register one another. Observation, here, means staying with rather than seeing through.

Although our methods use technical instruments, they do not aim to define or assess. Using swept-frequency capacitive sensing (SFCS) and an artificial neural network (ANN), we record micro-patterns of presence—contact zones formed through overlapping spatial interactions—without fixing them into discrete meanings. The resulting visuals are temporal records of contact, maintaining ambiguity and inviting reflection on what cannot be resolved.

Tracing without translation emphasizes relation and entanglement. Even when species do not share the same moment, their traces overlap and influence one another. The pond gathers these dispersed gestures into a subtle web of relation, showing that coexistence often unfolds indirectly. By avoiding labels or attribution, the system performs de-labeling—resisting human-centered interpretation and allowing relations to appear on their own terms. Water mediates this ecology, receiving every touch equally—whether from insect, leaf, or human hand—without hierarchy or judgment. Its neutrality embodies a quiet form of care that holds difference without fixing it.

“Observation without translation” thus becomes a practice of remaining with what is unresolved, framing design as reflective participation sustained by water’s quiet, mediating presence.

## 2 Sensus Pond System

### 2.1 Designing the pond as a sensing medium

This methodological stance takes shape through the design of the pond as a testbed that treats water not only as a site of observation but as a sensing medium. Rather than serving as a static backdrop, the pond functions as a dynamic field capable of registering subtle traces of contact. Water, though non-sentient, is materially responsive: it ripples, absorbs, and briefly holds disturbances caused by passing bodies. A landing insect, a bathing bird, a drinking cat, or a human hand—all leave small, temporary disruptions with the potential for quiet registration.

Through design, we aim to notice and record these short-lived interactions as temporal traces without fixing them into stable forms or assigned meanings. The pond becomes an active sensing medium, registering multispecies presence through continuous surface transformations. To support this approach, we avoid external sensing systems such as cameras or radar that observe from outside. Instead, we use SFCS embedded beneath the surface, detecting variations in contact as shifts in local conductivity. By working with water’s materiality from within, the system registers presence without translation, emphasizing relational and temporal dynamics as they unfold. It maintains indeterminacy as a condition for design, staying with rather than reducing more-than-human traces.

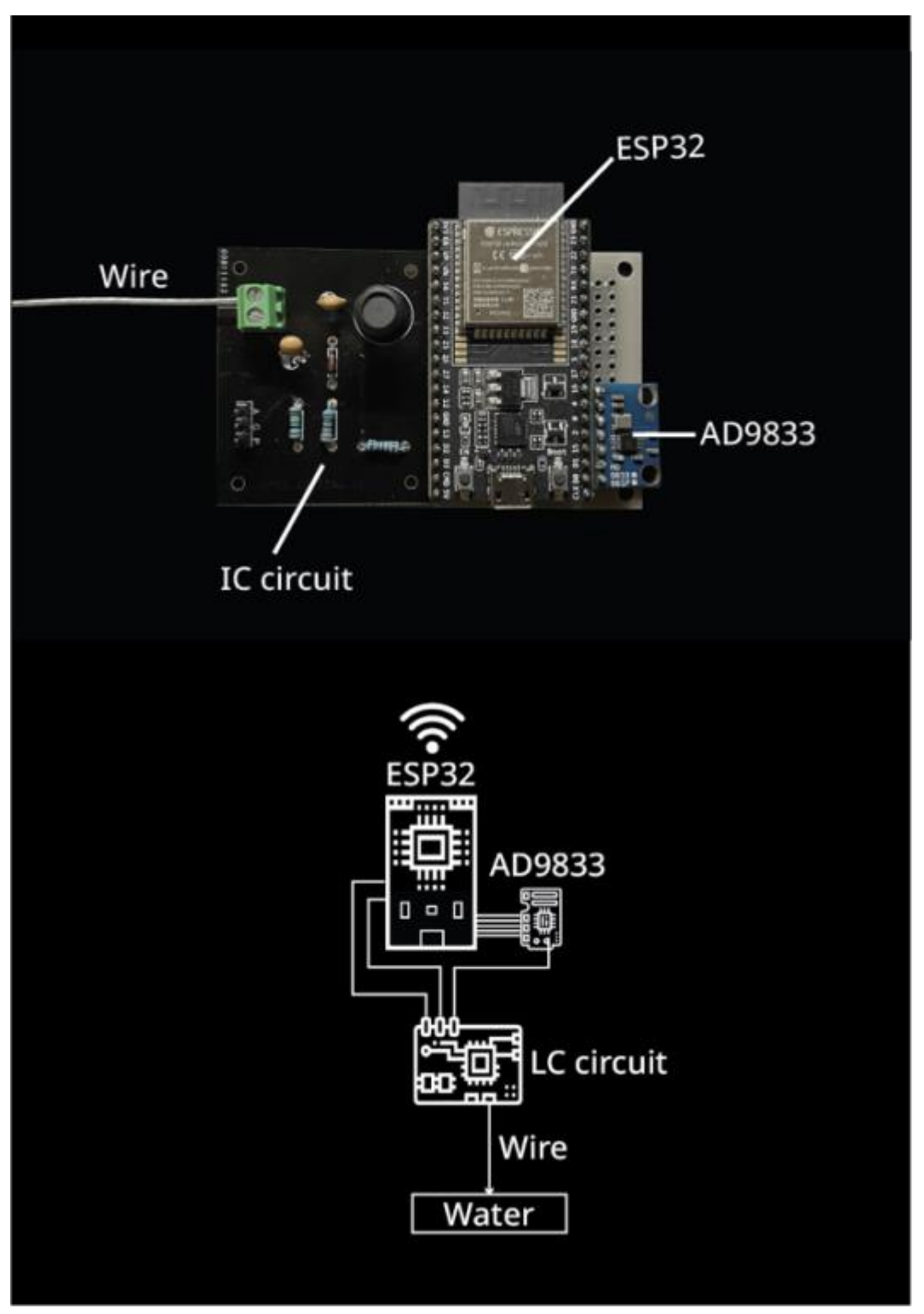

*Figure 3. SFCS unit configuration. It consists of an ESP32, and AD9833 wave generator and custom LC circuit.*

*Figure 4. SFCS signal changes with contact distance from sensing wire.*

## 2.2 Water sensing with SFCS and AI

We explored embedded sensing approaches that directly engage with the materiality of water. While sonar and electromagnetic methods effectively exploit water's physical properties (Hou et al., 2023; Jia et al., 2025), they rely on bulky hardware and complex infrastructure, making them unsuitable for lightweight, materially integrated systems. To address this, we adopted swept-frequency capacitive sensing (SFCS)—introduced in the Touché project (Sato et al., 2012)—which detects subtle interactions by measuring dielectric variations across frequencies. Requiring only a single wire embedded in the medium, SFCS registers contact directly through the material itself.

Its flexibility makes SFCS a promising foundation for embedded, material-based interfaces. It has been applied to sensing in water (Sato et al., 2012), plants (Poupyrev et al., 2012), textiles (Gong et al., 2021), and flexible substrates (Sakura & Kakehi, 2025). Building on these precedents, our work extends SFCS to larger water bodies. Through iterative design and technical development, we created a scalable sensing network composed of multiple SFCS nodes tuned for different ranges. Embedded within the pond, the network leverages water's conductive properties to detect contact trajectories. An artificial neural network (ANN) processes node data to estimate two-dimensional contact positions, enabling fine-grained tracing of surface interactions.

### 2.2.1 SFCS node development

As part of our water sensing system, we developed a custom swept-frequency capacitive sensing (SFCS) unit. Figure 3 shows its configuration: an ESP32 microcontroller, an AD9833 waveform generator, and a custom LC circuit designed for high-frequency signal processing. Inspired by Mads

Hobye's Arduino-based SFCS prototype (Hobye, 2012), our design extends it into a scalable, wireless architecture. By integrating the ESP32, we replaced the original wired serial setup with wireless UDP communication, greatly reducing wiring complexity and enabling flexible deployment across larger water bodies.

The SFCS operates by generating waveforms that sweep through a frequency range and measuring analog responses at each point. Because the ESP32 lacks the precise hardware timers required for stable high-frequency PWM generation, we incorporated the AD9833—a compact programmable signal generator capable of producing sine, triangular, and square waveforms across a broad spectrum. Through the SPI interface, the ESP32 controls the AD9833 to output square waves between 125 kHz and 10 MHz, which are fed into a custom LC circuit. Composed of an inductor and capacitors, this resonant circuit detects subtle dielectric variations within the medium, allowing the system to sense water's changing electrical properties and translate them into meaningful patterns of interaction.

To capture tactile interactions within the water body, the SFCS interface connects through a single wire embedded directly into the medium, creating a minimal yet materially entangled configuration (Fig. 4). This setup enables the unit to sense dielectric perturbations caused by external contact. As shown in Figure 3, when a hand touches the water surface at varying distances from the wire, the ESP32 records distinct spectral responses across a 10 MHz–125 kHz sweep, forming an 80-dimensional matrix. These spectra exhibit distance-dependent interference patterns, most evident in the mid-to-low frequency range, showing how proximity alters signal magnitude and shape. Acting as both a localized sensor and a node in a distributed network, the SFCS treats water not as an object to be sensed but as the sensing medium itself. In the next section, we extend this principle to a multi-node system using an ANN to infer two-dimensional contact positions from aggregated spectral data.

### 2.2.2 Sensing network with ANN-based trajectory analysis

Building on our SFCS unit configuration, we deployed a multi-node sensing network to enable spatial localization and temporal tracking of interactions across the water surface. The number of SFCS units can be flexibly adjusted depending on the sensing area. Each unit operates autonomously with wireless communication, forming a distributed array of embedded sensors within the same water body. Figure 5 shows an example setup with four SFCS units arranged around a pond. When a human or animal interacts with the water, the event is detected simultaneously by multiple sensor nodes through the shared medium. Each unit transmits its 80-dimensional spectral response via UDP to a computer, where the data streams are synchronized and aggregated in real time. The combined data are processed by an ANN-based regression model trained to estimate two-dimensional contact locations on the surface and to record their temporal sequence for trajectory tracking. Currently, the system detects only one contact at a time; simultaneous interactions by multiple entities cannot yet be separated, a limitation to be addressed in future iterations.

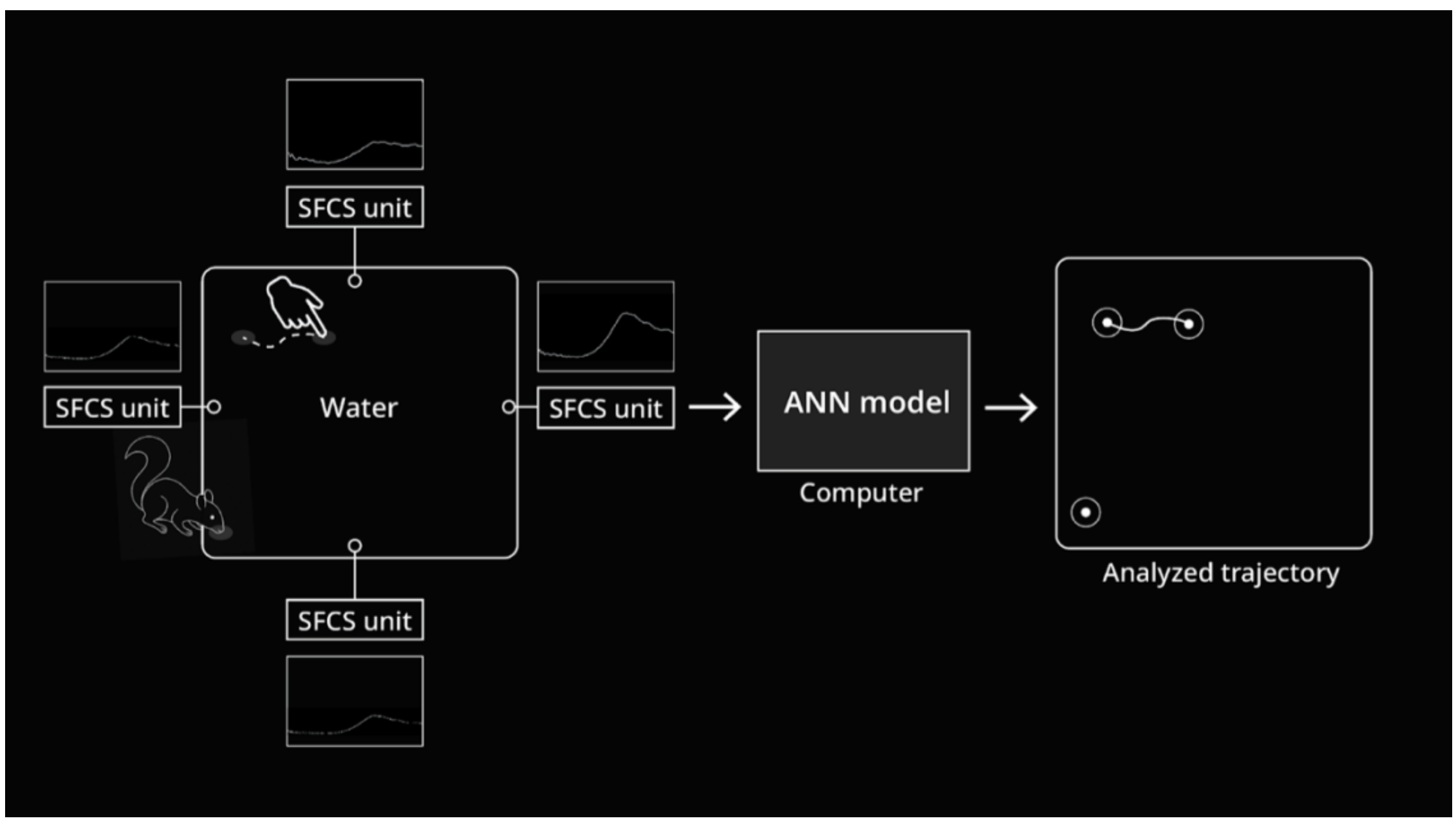

*Figure 5. System workflow of an example setup comprising four SFCS units.*

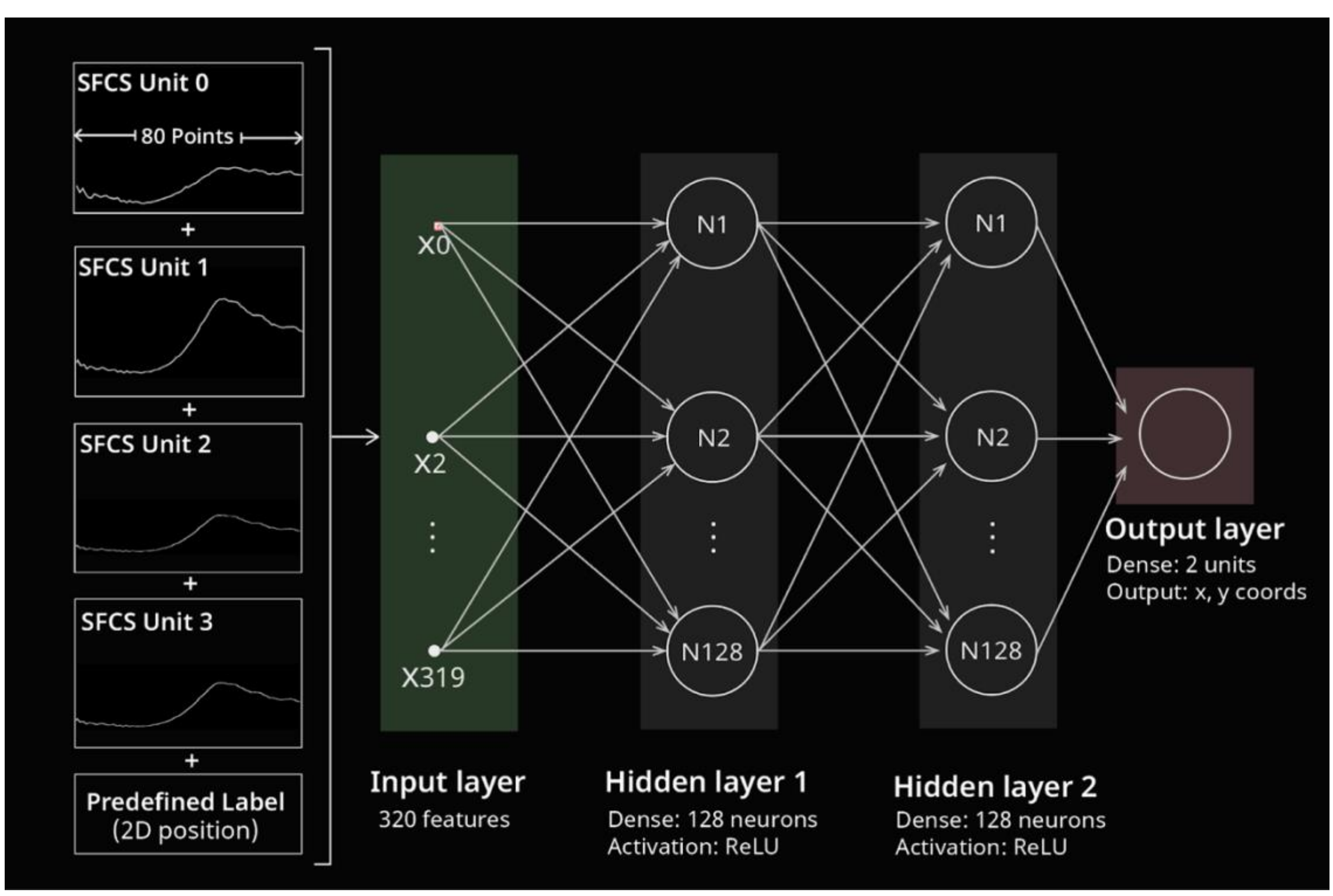

*Figure 6. Architecture of the ANN regression model implemented in the multi-node sensing network.*

We trained the ANN regression model using supervised learning. Figure 6 shows the four-layer architecture. The input layer receives a 320-dimensional feature vector formed by concatenating the 80-dimensional spectral responses from each of the four SFCS units. Ground-truth labels are the predefined (x, y) coordinates of contact points collected during calibration.

To build the dataset, we sampled five contact positions, each recorded 50 times, for a total of 250 samples. Positions included points near each SFCS unit and one central location. All input features were normalized to the [0, 1] range according to the 10-bit (1023) resolution of the ESP32 ADC. An 80/20 split divided the data into training and testing sets to evaluate model generalization.

The ANN was implemented in TensorFlow's Keras API with two fully connected hidden layers of 128 neurons using ReLU activation. The output layer contained two neurons corresponding to the predicted x and y coordinates. The model was trained with the Adam optimizer and mean squared error loss over 400 epochs and a batch size of 32. After training, the network mapped multi-node spectral input data to spatial contact locations and trajectories in real time, enabling continuous tracking of interactions on the water surface.

## 3 Traces of Contacts in Sensus Pond

Building on the conceptual grounding of tracing and observation without translation, we turned toward material encounters with *Sensus Pond*—deploying it in two distinct environments to explore how context shapes the emergence of more-than-human traces. These deployments were not designed as controlled experiments, but as open-ended inquiries into how gestures, presences, and environmental forces coalesce through water as a sensing medium. By tracing without translating, we sought to foreground relations and entanglements rather than discrete events, allowing the pond to reveal how contact unfolds across space and time.

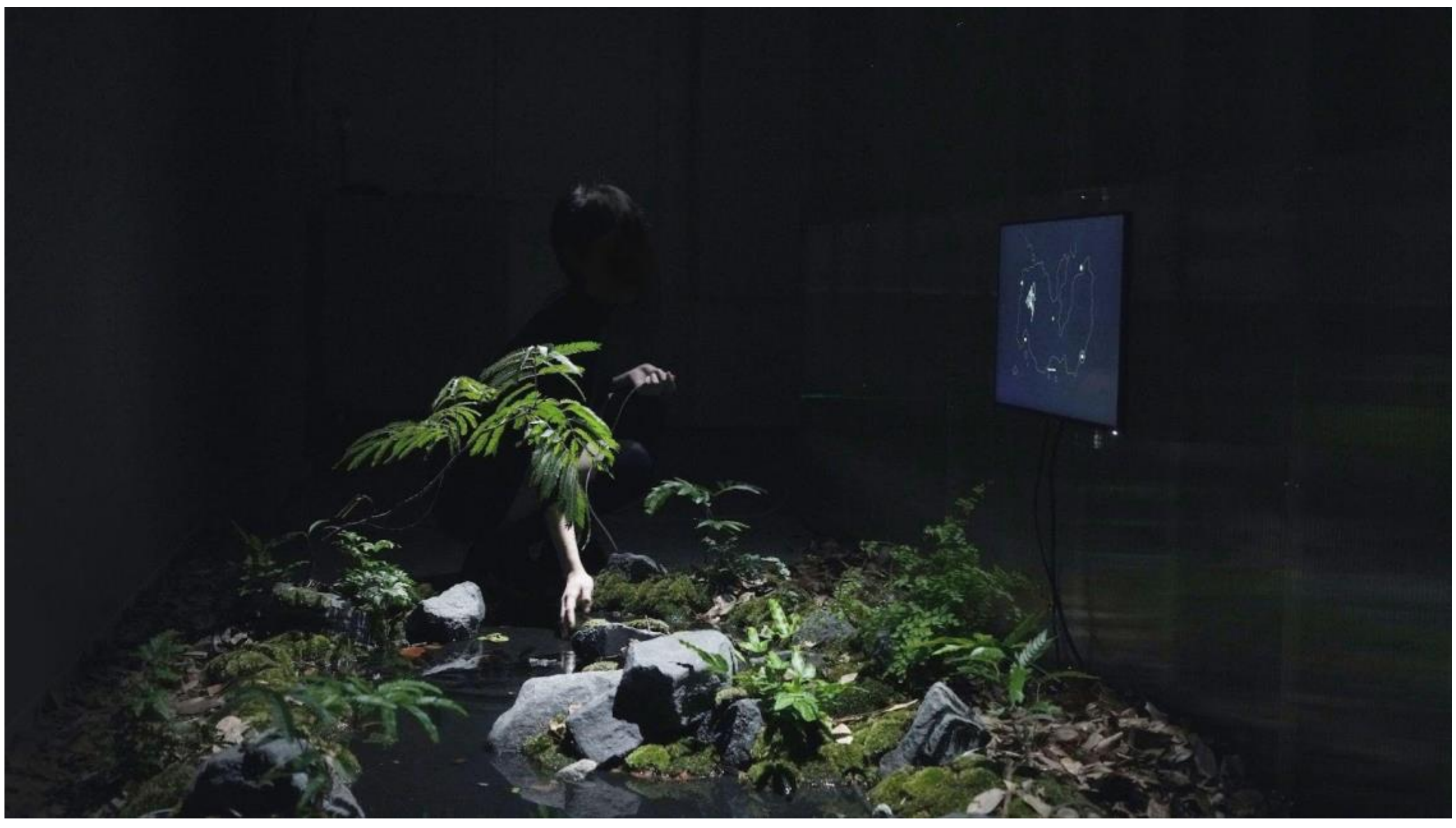

*Figure 7. Sensus Pond in the indoor environment.*

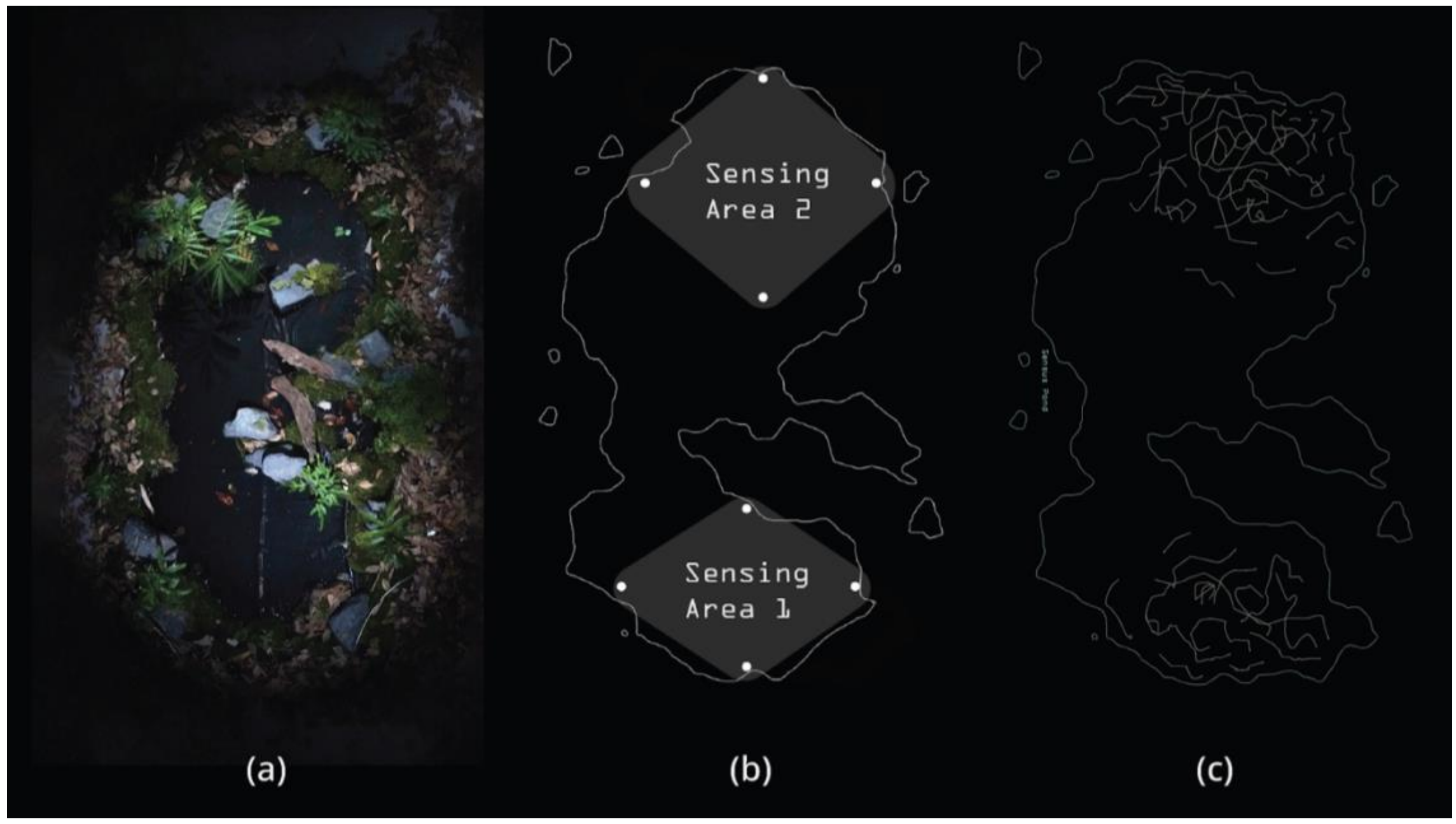

*Figure 8. Sensus Pond configuration in an indoor space: (a) Overhead view of the installed setup, (b) Arrangement of two designated sensing zones, (c) Visualization of cumulative interaction trajectories.*

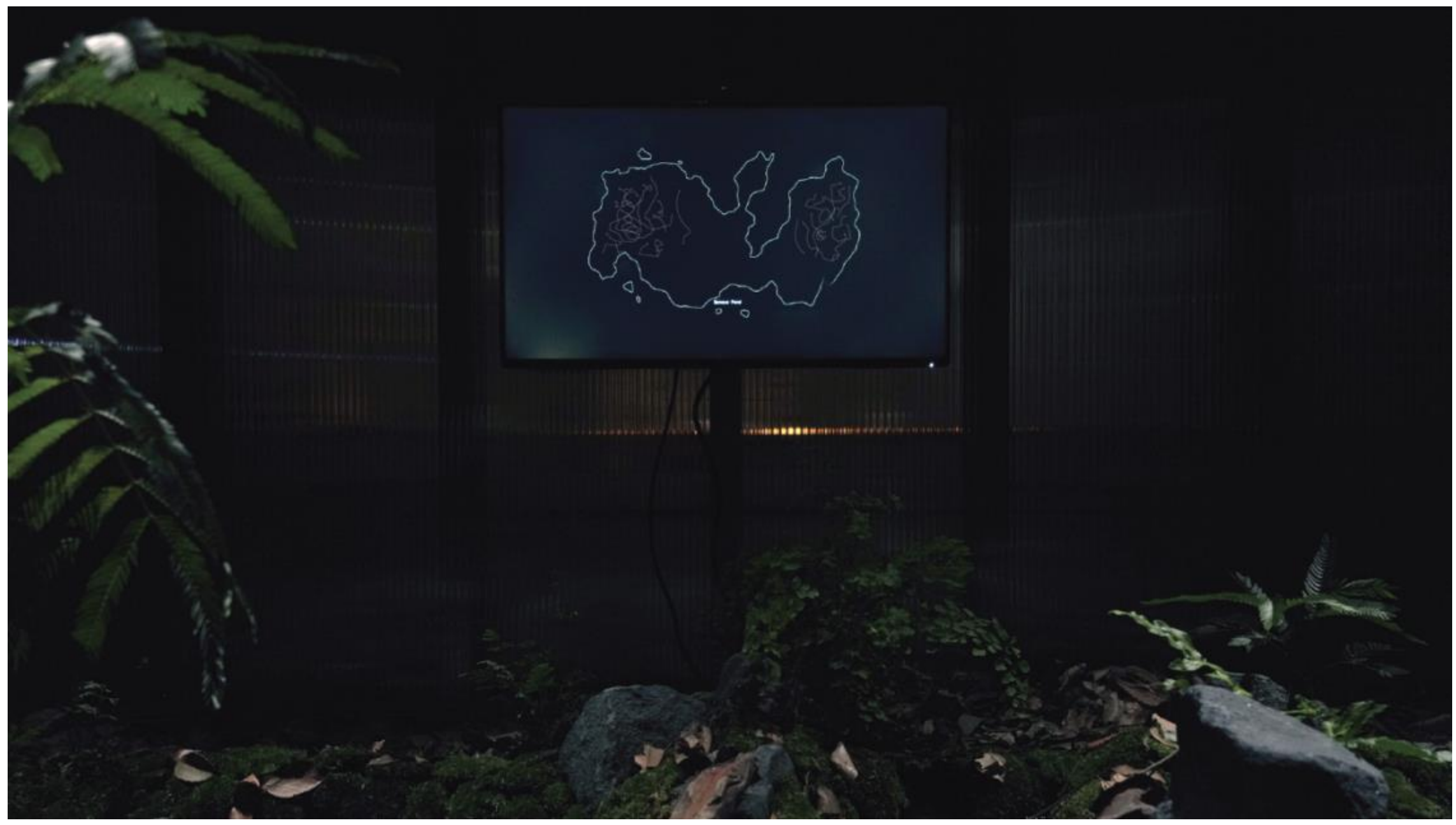

*Figure 9. Visualization of cumulative interaction trajectories on a wall-mounted display.*

## 3.1 Indoor trial

Our first installation took place in an indoor setting (Fig. 7), where we constructed a 2 m² water surface embedded with two sensing zones, each composed of four swept-frequency capacitive sensing (SFCS) units (Fig. 8a and b). A wall-mounted display visualized cumulative traces in real time, updating as interactions occurred across the water (Fig. 9). Over a three-day period, the visualization—implemented in openFrameworks and informed by ANN-based regression from Python—rendered the gradual accumulation and dissipation of human gestures (Fig. 8c).

These visualizations consist of trajectories formed by connecting individual data points recorded over time. The current form of the data appears as these simple linear traces—they were not intended to interpret or classify behavior. Rather, they reveal densities of contact: the quiet accumulation of touch, pause, and movement, forming a temporal cartography of spatial entanglement. What emerges is not a record of individuals, but layered trajectories of coexistence—the pond remembering, forgetting, and reconfiguring contact through its own dynamics of surface tension and reflection.

In this indoor configuration, tracing became a way of reading relation through residue. Each ripple and fluctuation left behind a soft imprint—a momentary correspondence between water and body. Observation was less about knowing who caused what, and more about sensing how the pond itself participates in the act of remembering—how matter keeps trace. In addition to the current form, we also plan to extend the duration of data accumulation to capture longer temporal rhythms and to explore new modes of visualization beyond the current representations, seeking forms that further reveal the subtle entanglements between water, body, and time.

## 3.2 Semi-natural deployment

To extend this inquiry beyond human presence, we relocated Sensus Pond to a semi-natural setting: a 4 × 6 m backyard bordered by shrubs and neighboring buildings (Fig. 10). The new pond, about 1 m in diameter and 20 cm deep, was left uncovered—open to lizards, insects, birds, neighborhood cats, and occasional human visitors.

Unlike the controlled indoor setup, this environment introduced shifting light, changing weather, and multispecies rhythms. Here, water acted as a neutral mediator, absorbing every contact equally—a bird's landing, a falling leaf, a cat's tentative paw. Its surface gathered these gestures into overlapping traces of coexistence, showing that species need not share the same moment to be entangled. The system avoided labeling or categorizing each trace, allowing patterns of relation to emerge through quiet persistence. Visual outputs were not explanations but temporal surfaces of ambiguity, supporting slower and more open observation.

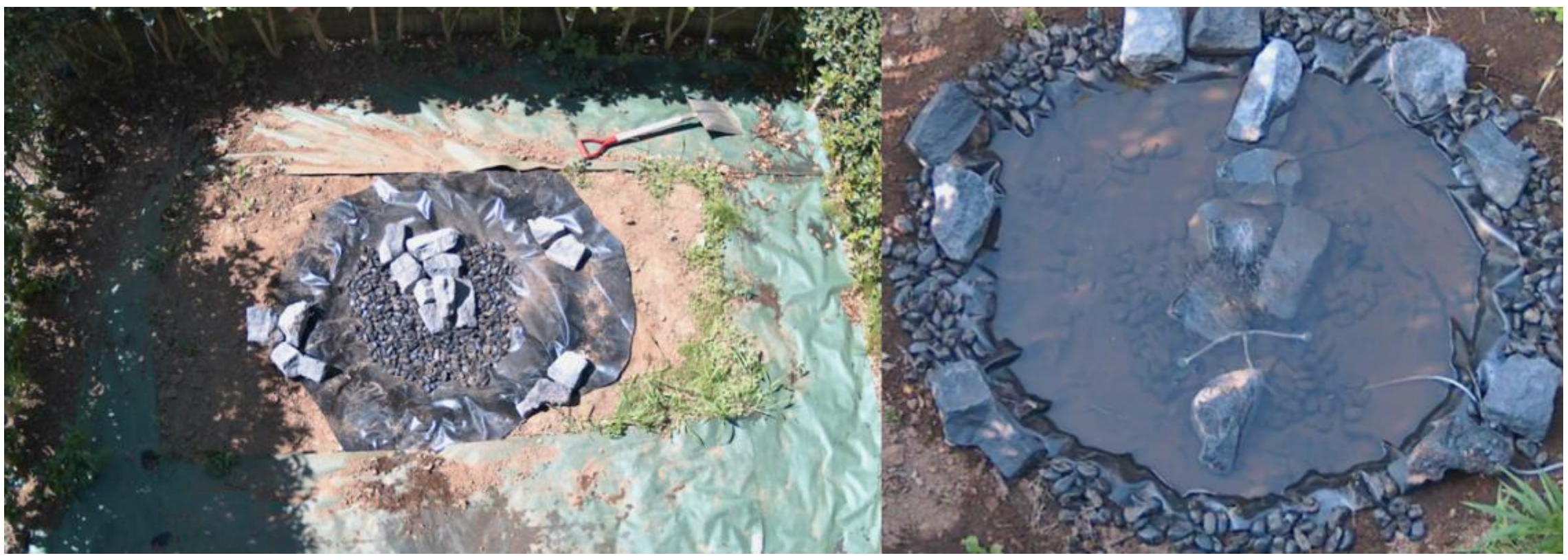

*Figure 10. Process of building the Sensus Pond in a backyard.*

The ongoing outdoor experiments are being conducted as part of a long-term study. Over time, we expect the pond to gradually influence its surroundings—a neglected patch of soil becoming a zone of shared attention, a small commons where human and nonhuman rhythms meet. Future work will include more systematic analysis of these observations, as well as comparative and quantitative evaluation of multispecies interaction patterns over extended periods of time. Through these analyses, we aim to understand how cumulative traces reveal rhythms of cohabitation—not as individual acts, but as ongoing entanglement across species and seasons.

## 4 Conclusion

This pictorial presents Sensus Pond, which reconfigures water from a static element into an active sensing medium. Through embedded design and site-specific deployment, the pond offers another way of observing more-than-human traces that resist classification, representation, and fixed meaning. It demonstrates tracing as a methodological stance of attunement and non-translation, shifting from intervention to observation. The system registers direct contact—light taps, gliding touches, brief immersions—each leaving a short trace that accumulates over time.

The pond combines SFCS units with ANN to estimate spatial interaction patterns. This setup supports situated, material sensing that records contact without interpretation. Rather than isolating events, it focuses on their spatial and temporal buildup, as the surface gradually records overlapping presences. These traces are not categorized or evaluated but approached as local impressions that remain contingent and open-ended.

Together, these practices enact observation without translation—an approach that resists legibility and accepts ambiguity as part of design inquiry. The pond becomes a medium for exploring shared habitation and for reconsidering how sensing is framed and situated within design as ongoing correspondence with ecological systems.

### Acknowledgements

This research was supported by The Univeristy of Tokyo - NTT Sustainable Well-being Social Collaboration Initiative. We also thank the help from xlab members.

## References

Wolfe, C. (2010). What is posthumanism? (Vol. 8). U of Minnesota Press. https://doi.org/10.1086/670305

Barad, K. (2003). Posthumanist performativity: Toward an understanding of how matter comes to matter. Signs: Journal of women in culture and society, 28(3), 801-831. https://doi.org/10.1086/345321

Haraway, D. J. (2016). Staying with the trouble: Making kin in the Chthulucene. In Staying with the Trouble. Duke University Press. https://doi.org/10.1515/9780822373780

Giaccardi, E., Redström, J., & Nicenboim, I. (2025). The making (s) of more-than-human design: introduction to the special issue on more-than-human design and HCI. Human–Computer Interaction, 40(1-4), 1-16. https://doi.org/10.1080/07370024.2024.2353357

Rosén, A. P., Normark, M., & Wiberg, M. (2022, October). Noticing the environment–A design ethnography of urban farming. In Nordic Human-Computer Interaction Conference (pp. 1-13). https://doi.org/10.1145/3546155.3546659

Chang, M., Shen, C., Maheshwari, A., Danielescu, A., & Yao, L. (2022, June). Patterns and opportunities for the design of human-plant interaction. In Proceedings of the 2022 ACM Designing Interactive Systems Conference (pp. 925-948). https://doi.org/10.1145/3532106.3533555

Hu, Y., Fol, C. R., Chou, C., Griess, V. C., & Kakehi, Y. (2024, May). Immersive Flora: Re-Engaging with the Forest through the Visualisation of Plant-Environment Interactions in Virtual Reality. In Extended Abstracts of the CHI Conference on Human Factors in Computing Systems (pp. 1-6). https://doi.org/10.1145/3613905.3648675

Liu, J., Byrne, D., & Devendorf, L. (2018, April). Design for collaborative survival: An inquiry into human-fungi relationships. In Proceedings of the 2018 CHI Conference on Human Factors in Computing Systems. https://doi.org/10.1145/3173574.3173614

Lu, J., & Lopes, P. (2022, October 29). Integrating living organisms in devices to implement care-based interactions. Proceedings of the 35th Annual ACM Symposium on User Interface Software and Technology. UIST ’22: The 35th Annual ACM Symposium on User Interface Software and Technology, Bend OR USA. https://doi.org/10.1145/3526113.3545629

Ofer, N., & Alistar, M. (2024, July). Tracing as a strategy for orienting to nonhuman perspectives. In Proceedings of the 2024 ACM Designing Interactive Systems Conference (pp. 1087-1100). https://doi.org/10.1145/3643834.3661621

Pataranutaporn, P., Vujic, A., Kong, D. S., Maes, P., & Sra, M. (2020, March). Living bits: Opportunities and challenges for integrating living microorganisms in human-computer interaction. In Proceedings of the augmented humans international conference (pp. 1-12). https://doi.org/10.1145/3384657.3384783

Zhou, J., Doubrovski, Z., Giaccardi, E., & Karana, E. (2024, May). Living with Cyanobacteria: Exploring Materiality in Caring for Microbes in Everyday Life. In Proceedings of the 2024 CHI Conference on Human Factors in Computing Systems (pp. 1-20). https://doi.org/10.1145/3613904.3642039

Bell, F., Chow, D., Choi, H., & Alistar, M. (2023, February). Scoby breastplate: Slowly growing a microbial interface. In Proceedings of the seventeenth international conference on tangible, embedded, and embodied interaction (pp. 1-15). https://doi.org/10.1145/3569009.3572805

Ikeya, Y., Wakkary, R., & Barati, B. (2023, July). Metamorphonic: A reflective design inquiry into human-silkworm relationship. In Proceedings of the 2023 ACM Designing Interactive Systems Conference (pp. 808-819). https://doi.org/10.1145/3563657.3596053

Wakkary, R., Oogjes, D., Sakib, N., & Behzad, A. (2023, July). Turner Boxes and Bees: From Ambivalence to Diffraction. In Proceedings of the 2023 ACM Designing Interactive Systems Conference (pp. 790-807). https://doi.org/10.1145/3563657.3596081

Wakkary, R. (2021). Things we could design: For more than human-centered worlds. MIT press. https://doi.org/10.7551/mitpress/13649.001.0001

Hui, Y. (2021). Art and cosmotechnics. U of Minnesota Press. https://doi.org/10.5749/j.ctv1qgnq42

Wakkary, R., Oogjes, D., Tomico, O., Sakib, N., & Kökel, E. (2025, April). Backyard Practices: A Liminal Approach to Designing in More-than-Human Worlds. In Proceedings of the 2025 CHI Conference on Human Factors in Computing Systems (pp. 1-18). https://doi.org/10.1145/3706598.3713291

Offenhuber, D. (2019). Data by proxy—material traces as autographic visualizations. IEEE transactions on visualization and computer graphics, 26(1), 98-108. https://doi.org/10.1109/TVCG.2019.2934788

Hou, M., Wu, H., Peng, J., & Li, K. (2023). Long-range and high-precision localization method for underwater bionic positioning system based on joint active–passive electrolocation. Scientific reports, 13(1), 21475. https://doi.org/10.1038/s41598-023-48957-x

Jia, L., Zhang, G., Liu, Y., Bai, Z., Geng, Y., Wu, Y., ... & Zhang, W. (2025). Sonar buoy active detection and localization for underwater targets using high-level sound sources and MEMS hydrophone. Measurement, 241, 115740. https://doi.org/10.1016/j.measurement.2024.115740

Sato, M., Poupyrev, I., & Harrison, C. (2012, May). Touché: enhancing touch interaction on humans, screens, liquids, and everyday objects. In Proceedings of the SIGCHI Conference on Human Factors in Computing Systems (pp. 483-492). https://doi.org/10.1145/2207676.2207743

Poupyrev, I., Schoessler, P., Loh, J., & Sato, M. (2012). Botanicus Interacticus: interactive plants technology. In ACM SIGGRAPH 2012 Emerging Technologies (pp. 1-1). https://doi.org/10.1145/2343456.2343460

Sakura, R., & Kakehi, Y. (2025, April). A 3D-Printed Touch Sensor with a Single-Stroke Conductive Path. In Proceedings of the Extended Abstracts of the CHI Conference on Human Factors in Computing Systems (pp. 1-6). https://doi.org/10.1145/3706599.372002

Gong, H., Cui, Z., Wang, Y., Shen, C., Zhang, D., & Luo, S. (2021, May). EGlove: Designing interactive fabric sensor for enhancing contact-based interactions. In Extended Abstracts of the 2021 CHI Conference on Human Factors in Computing Systems (pp. 1-7). https://doi.org/10.1145/3411763.345182

Hobye, M. (2012). Touche for Arduino: Advanced Touch Sensing. https://www.instructables.com/Touche-for-Arduino-Advanced-touch-sensing/

*We utilized generative AI for language editing, and the concept sketch of the Sensus Pond environment was created with image generation AI.